\documentclass[aps,pra,superscriptaddress,amsmath,amssymb,preprintnumbers,showpacs,twocolumn]{revtex4}
\usepackage{amssymb}
\usepackage{graphicx}
\usepackage{bm}
\usepackage{color}

\begin{document}

\title{\small Quantum Phase Transitions and Heat Capacity in a two-atoms Bose-Hubbard Model}

\author{B. Leggio}
\address{Dipartimento di Fisica, Universit\`{a} di Palermo, Via Archirafi 36, 90123 Palermo,
Italy}

\author{A. Napoli}
\address{Dipartimento di Fisica, Universit\`{a} di Palermo, Via Archirafi 36, 90123 Palermo,
Italy}

\author{A. Messina}
\address{Dipartimento di Fisica, Universit\`{a} di Palermo, Via Archirafi 36, 90123 Palermo,
Italy}

\newcommand{\ket}[1]{\displaystyle{|#1\rangle}}
\newcommand{\bra}[1]{\displaystyle{\langle #1|}}

\begin{abstract}
We show that a two-atoms Bose-Hubbard model exhibits three different phases in the behavior of thermal entanglement in its parameter space. These phases are demonstrated to be traceable back to the existence of quantum phase transitions in the same system. Significant similarities between the behaviors of thermal entanglement and heat capacity in the parameter space are brought to light thus allowing to interpret the occurrence and the meaning of all these three phases.
\end{abstract}
\maketitle

\section{Introduction}
A Quantum Phase Transition (QPT) is a phenomenon \cite{sac} consisting on a qualitative change of the ground state of a quantum system as one or more of
its physical parameters are varied \cite{rev, wa, expqpt, bon}. These phase transitions are
genuinely quantum mechanical effects directly stemming from
quantum fluctuations. Differently from thermal phase transitions
indeed, they occur at zero temperature and manifest themselves
through the non-analytical behavior of the ground state energy at
a transition point \cite{wan, zh, dhar} in the parameter space of the physical system, around which some selected physical properties of the system experience significant, sometimes sharp, modifications \cite{ost, shi}. For instance recent investigations on some systems
undergoing quantum phase transitions have demonstrated that the transition is related to a marked change in
the entanglement \cite{wu}. Depending on the physical system considered, entanglement could have a peak,
show discontinuous behavior or also manifest diverging derivatives
at the critical points \cite{osb, vid}. It is worth noting however that, generally speaking, such transitions are accompanied by correlations whose nature is not clear enough \cite{dil, sma, or}.
Since the degree of entanglement getting established between different components of a
system cannot be measured in laboratory, the ability of finding
easily measurable observables that can be used as entanglement
witnesses represents an important challenge. This paper investigates on a
simple but not trivial system consisting of few atoms in an
optical lattice, bringing to light the existence of QPTs and their
connection with the thermal entanglement characterizing the
system. After that we find our main result, that is a link between negativity and heat capacity behaviors in the parameter space, leading to a deeper understanding of the origin and the occurrence of a QPT.\\
This work is organized as follows: in Section II we discuss and analyze the Hamiltonian model describing our physical system evaluating as well all quantities of interest. The main results of this paper are presented in Section III. Comments and conclusions are drawn in Section IV.

\section{Hamiltonian Model}
Very recently the Hubbard-type Hamiltonian \cite{blo} ($\hbar=1$)
\begin{eqnarray} \label{hambh}
H_{BH}=&-t\sum_{<ij>,\sigma}\big( a_{i\sigma}^{\dag}a_{j\sigma} + a_{j\sigma}^{\dag}a_{i\sigma} \big)+\epsilon\sum_{i}\hat{n}_{i}\nonumber \\
&+\frac{U_{0}}{2}\sum_{i}\hat{n}_{i}(\hat{n}_{i}-1)+\sum_{i}\frac{U_{2}}{2}\Big((\mathbf{S}_{tot}^{i})^{2}-2\hat{n}_{i} \Big) \nonumber \\
\end{eqnarray}
describing spin-1 atoms in an optical lattice, has attracted much attention \cite{rizzi, bai, chn, gen} thanks to the renewed interest in spin gases stemming from new experimental techniques allowing a detailed control over multi-particle states \cite{zz}. In eq. (\ref{hambh})
$a_{i\sigma}^{\dag}$ is the creation operator of an atom in the
$\emph{i}-th$ lattice site with spin projection over a fixed
quantization axis equal to $\sigma=(-1,0,1)$,
$\hat{n}_{i}=\sum_{\sigma}a_{i\sigma}^{\dag}a_{i\sigma}$ is the
total number of atoms on site $i$ and $\textbf{S}_{tot}^{i}$ is
the total spin operator of the $\emph{i}-th$ lattice site. The
first term of $H_{BH}$ describes spin symmetric tunneling between
nearest neighboring sites, the parameter $t$ being the tunneling
amplitude and the  $\sum_{<ij>}$  being extended over all pairs of
neighboring sites. The second term describes the Hubbard
repulsion between atoms, $U_0$ being the contact scattering
amplitude, whereas the term proportional to $U_2$ differentiates scattering channels with $\textbf{S}_{tot}$ equals
to $0$ or $2$, since scattering lengths for these two channels are
not the same \cite{projectors}.
Finally $\epsilon$ is the atom self energy which may be neglected in
what follows. \\

Let us assume that two different atoms are described by states
centered in different lattice sites and that the temperature is
low enough to consider the probability of finding an atom outside
the site on which its state is centered almost zero.

Under this hypothesis the tunneling term can be treated as a
perturbation and as a consequence, always at low temperature, the tunneling term in Hamiltonian (\ref{hambh}) can be replaced by an effective next-neighbors interaction \cite{hameff}. Tunneling processes induce effective pairwise interactions
between atoms on neighboring sites and the tunneling term in Hamiltonian (\ref{hambh}), perturbatively expanded up to second order in parameter $t$, may be substituted with the following effective one
\begin{equation} \label{hameff}
H_{t}^{eff}=\widetilde{\omega} J_{z}+ K_{0} + K_{1}\sum_{<ij>}(\textbf{S}_{i}\cdot\textbf{S}_{j})+K_{2}\sum_{<ij>}(\textbf{S}_{i}\cdot \textbf{S}_{j})^{2}
\end{equation}
in which an external magnetic field $\widetilde{\omega}$ has been added to the
system described above. Here $\textbf{S}_{i}$ represents the spin
operator of the particle in the \emph{i}-th site and
$\textbf{J}=\sum_{i}\textbf{S}_{i}$. The effective coupling
constants $K_{0}$, $K_{1}$, $K_{2}$ are related to the parameters
$t, U_0$ and $U_2$ appearing in the microscopic Hamiltonian (\ref{hambh}) as
follows
\begin{eqnarray}
&K_{0}=\frac{4t^{2}}{3(U_{0}+U_{2})}-\frac{4t^{2}}{3(U_{0}-2U_{2})} \label{k0} \\
&K_{1}=\frac{2t^{2}}{U_{0}+U_{2}} \label{k1} \\
&K_{2}=\frac{2t^{2}}{3(U_{0}+U_{2})}+\frac{4t^{2}}{3(U_{0}-2U_{2})} \label{k2}
\end{eqnarray}
and satisfy the simple equation $K_{0}=K_{1}-K_{2}$. In the deep Mott-insulator phase, where (\ref{hameff}) holds, the tunneling amplitude $t$ is of the order of some kHz but can be as well as small as a few Hz \cite{expbhqpt}.\\
We remark that the term (\ref{hameff}) is an effective one describing the
physics of hopping between lattice sites in terms of interactions between spin operators of the atoms. Two interaction terms appear,
a linear and a quadratic one, and their origin can be
qualitatively understood. Let us indeed consider atoms whose states
are centered in different sites. After the hopping of one of these
atoms towards one of its neighboring sites there will be a
potential well containing two particles, thus allowing them to
interact by s-wave scattering. On the other hand two spin-1
particles scattering inside a potential well may result in two
different final states with total spin equal to $0$ or $2$ (states
with total spin equal to $1$ are forbidden for symmetry reasons).
Thus, in order to properly describe these two different situations
in terms of spin-spin interaction, one needs to express the
projectors $P_{0}$ and $P_{2}$ ($P_{s}$ being the operator
projecting the state of two atoms into a final state with total
spin equal to $s$) in terms of spin operators of the particles
involved. It is possible to show \cite{projectors} that this requires quadratic interactions together
with linear ones. In this paper we concentrate on the simple, but
not trivial, situation in which only two atoms are considered. With this further approximation one can look at $H_{t}^{eff}$ as the full Hamiltonian of the system.\\
Measuring the energy in units of $t$, $H_{t}^{eff}$ in the case of two atoms only can be cast in the following form
\begin{equation} \label{hamzl}
\frac{H_{t}^{eff}}{t}\equiv H=\omega J_{z}+\tau(\textbf{S}_{1}\cdot
\textbf{S}_{2})+\gamma(\textbf{S}_{1}\cdot \textbf{S}_{2})^{2}+r I
\end{equation}
where
\begin{eqnarray}
&\tau= K_{1} / t \nonumber \\
&\gamma= K_{2} / t \nonumber \\
&\omega=\widetilde{\omega} / t \nonumber \\
&r=\tau-\gamma
\end{eqnarray}
and $I$ is the identity operator in the 9-dimensional
Hilbert space of the system. We do not neglect the term proportional to identity since we want to investigate the system in its parameters space.\\
It is worth noting that investigations on few atoms models \cite{longhi, zieg, lin} are interesting in their own since under particular assumptions they describe small subsystems in larger lattices providing hints to understand their physical properties \cite{leg}. On the other hand considering few particles systems allows us to exactly diagonalize the Hamiltonian and to focus on the dynamically generated pairwise entanglement.

\subsection{Thermal entanglement}
To diagonalize the Hamiltonian given by eq. (\ref{hamzl}) it is convenient to write it in
the form
\begin{eqnarray} \label{hamzlrewr}
H&=\omega J_{z}+\frac{\tau}{2}(J^{2}-S_{1}^{2}-S_{2}^{2})+\frac{\gamma}{4}(J^{2}-S_{1}^{2}-S_{2}^{2})^{2}+ \nonumber \\
&+r I=\omega J_{z}+\frac{\tau}{2}(J^{2}-4I)+\frac{\gamma}{4}(J^{2}-4I)^{2}+r I \nonumber \\
\end{eqnarray}
where $\textbf{J}=\textbf{S}_1+\textbf{S}_2$.
$H$ is then diagonal in the
coupled basis $|jM\rangle$ of common eigenstates of $J^{2}$ and
$J_{z}$ with associated eigenvalues equal to $j(j+1)$ and $M$
respectively. The energy eigenvalues are
\begin{equation}\label{eigenvalues}
E_{jM} = \omega M + \frac{\tau}{2}(j(j+1)-2)+\frac{\gamma}{4}\big[(j(j+1)-4)^{2}-4\big]
\end{equation}
Let us suppose that our system is in a thermal state at $T$
temperature. The density matrix describing the two spin 1 atoms in
the optical lattice, can be thus written as
\begin{equation} \label{thermalstatezl}
\rho^{T}=\frac{1}{Z}e^{-\beta H}
\end{equation}
where $\beta = 1 / T$ $(k_B=1)$ and $Z =
Tr(e^{-\beta H})$.\\
$Z$ can be exactly given in closed form as follows
\begin{eqnarray} \label{partfunctzl}
Z=&e^{-\beta\tau}\Big[2\cosh{\beta\tau}\big(1+2\cosh{\beta\omega}\big) + \nonumber \\
&+2e^{-\beta\tau}\cosh{2\beta\omega} + e^{-\beta(3\gamma - 2\tau)} \Big]
\end{eqnarray}
In order to quantify thermal entanglement in the state
(\ref{thermalstatezl}) we use the well known negativity function
$N$ defined as \cite{peres}
\begin{equation}\label{negzl}
N=\frac{1}{2}\Big( \sum_{i=1}^{9}|\lambda_{i}|-1 \Big)
\end{equation}
where $\lambda_{i}$ are the eigenvalues of the matrix
$\sigma$, partial transpose of $\rho$ with respect to one of the two spins. Representing $\rho$ with respect to the ordered factorized basis $\Big\{|-11\rangle,|1-1\rangle,|-10\rangle,|01\rangle,|0-1\rangle,|10\rangle,|-1-1\rangle,|00\rangle,|11\rangle
\Big\}$, and transposing with respect to the spin labeled as 1 leads to
\begin{eqnarray}\label{rhotazl}
\sigma= \left(\begin{array}{cccccccccc}
R_{+} & 0 & 0 & 0 & 0 & 0 & 0 & 0 & 0 \\
0 & R_{+} & 0 & 0 & 0 & 0 & 0 & 0 & 0 \\
0 & 0 & P_{-} & Q_{-} & 0 & 0 & 0 & 0 & 0 \\
0 & 0 & Q_{-} & P_{+} & 0 & 0 & 0 & 0 & 0 \\
0 & 0 & 0 & 0 & P_{-} & Q_{-} & 0 & 0 & 0\\
0 & 0 & 0 & 0 & Q_{-} & P_{+} & 0 & 0 & 0 \\
0 & 0 & 0 & 0 & 0 & 0 & L_{-} & M_{-} & R_{-} \\
0 & 0 & 0 & 0 & 0 & 0 & M_{-} & Q_{+} & M_{+} \\
0 & 0 & 0 & 0 & 0 & 0 & R_{-} & M_{+} & L_{+}
\end{array}\right) \nonumber \\
\end{eqnarray}
where
\begin{eqnarray}
&L_{\pm}=\frac{1}{Z}e^{-2\beta(\tau\pm\omega)}\\
&M_{\pm}=-\frac{1}{Z}e^{-\beta(\tau\pm\omega)}\sinh{(\beta\tau)} \\
&P_{\pm}=\frac{1}{Z}e^{-\beta(\tau \pm \omega)}\cosh{\beta\tau} \\
&R_{\pm}=\frac{1}{6Z}e^{-\beta \tau}\big(e^{-\beta\tau}\pm 3e^{\beta\tau}+2e^{-\beta(3\gamma -2\tau)}\big) \\
&Q_{\pm}=\frac{1}{3Z}e^{-\beta\tau}\Big(\frac{3\pm1}{2} e^{-\beta\tau}\pm e^{-\beta(3\gamma -2\tau)} \Big)
\end{eqnarray}
Six of the nine eigenvalues can immediately be derived in the form
\begin{eqnarray}
&\lambda_{1}=\lambda_{2}=R_{+} \nonumber \\
&\lambda_{3}=\lambda_{5}=\frac{1}{2}\Big( P_{+}+P_{-}-\sqrt{\big( P_{+}-P_{-} \big)^{2}+4Q_{-}^{2}} \Big) \\
&\lambda_{4}=\lambda_{6}=\frac{1}{2}\Big( P_{+}+P_{-}+\sqrt{\big( P_{+}-P_{-} \big)^{2}+4Q_{-}^{2}} \Big) \nonumber
\end{eqnarray}
while the last three eigenvalues can be obtained solving the secular equation associated to the $3\times 3$ block
\begin{eqnarray}
B=\left(\begin{array}{ccc}
L_{-} & M_{-} & R_{-} \\
M_{-} & Q_{+} & M_{+} \\
R_{-} & M_{+} & L_{+} \\
\end{array}\right)
\end{eqnarray}
Since the algebraic expressions of these last three eigenvalues of $\sigma$ do not exhibit features deserving special attention we do not give them explicitly.\\
The knowledge of the eigenvalues $\lambda_{i}$ of $\sigma$, as functions of both temperature and the three model parameters, allows us to investigate, at a given temperature, the behavior of negativity in the parameter space. Recently the role of  the nonlinear effective coupling between the
two atoms on the thermal entanglement has been qualitatively
investigated. In detail it has been found that
the quadratic interaction term favors the thermal entanglement
that on the other hand exists only for absolute values of the quadratic
parameter $\gamma <0$ larger than a critical value $\gamma_c$ \cite{zhangli}. In
presence of an external magnetic field ($\omega\neq 0$) the system
shows two different phases, one correspondent to $N=1$ and the
other one to  $N=0$, and the value of $\gamma_c$ at which the
phase transition takes place is dependent on $\omega$. These
results have anyway been obtained neglecting from the very beginning the
linear interaction term whose intensity is measured by $\tau$ in Hamiltonian model (\ref{hamzl}). In
what follows we investigate on the role played by the
linear term. In particular we wonder whether there is competition between the linear and the quadratic
effective interaction terms or if they
cooperate in establishing entanglement.

\section{Results}
\begin{figure}[h]
\begin{center}
\includegraphics[width=250pt]{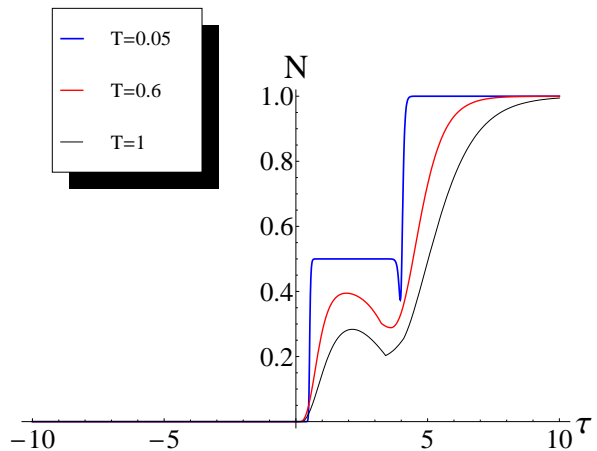}
\end{center}
\caption{System negativity plotted against $\tau$ when $\gamma =\omega =1$. Energy is measured in units of $t$ and $k_{B}=1$}
\label{negzl1}
\end{figure}

The negativity $N$ against the linear interaction parameter $\tau$, for $\omega$ and $\gamma$ fixed, is reported in Fig.\ref{negzl1} in correspondence to three values of temperature $T$, namely $T=0.05$, $T=0.6$ and $T=1$. The addition of the term proportional to $\tau$ to the model of Ref. \cite{zhangli} is at the origin of an intermediate phase (plateau of the negativity, $N=\frac{1}{2}$) for $\omega$ and $\gamma$ equal to 1, when $\tau$ runs around 2. $N$ undergoes indeed two phase transitions (vertical lines) differently from what happens when $\tau=0$ where only one phase transition is present (cf. Fig.1 of Ref. \cite{zhangli})
It is easy to interpret the occurrence of the intermediate phase at $N=\frac{1}{2}$ as due to the magnetic field assisted competition between the quadratic term and the linear one when both $\gamma$ and $\tau$ are positive. To appreciate this point it is enough to note that the two-atom system, when $\tau >0$, tends to minimize its total energy assuming $j=0$ (maximally entangled state), whilst it tends to maximize both $j$ and $|M|$ (factorized state) when $\gamma >0$ in presence of a magnetic field.\\
Figure 1 shows that the two phase transitions are relatively sharper in correspondence to the smallest value of the temperature $T$. Investigating on the behavior of the negativity $N$ when the
temperature $T$ goes toward zero, it is possible to demonstrate
that the smooth phase transitions shown in Fig.1, become indeed jumps. At the light of recent papers \cite{zh,zvy}, we wonder whether this behavior reflects the existence of QPTs for the system, in this way taking part to the topical debate about the connection which might exist between entanglement and QPTs. It has been proved
for example that in a class of spin systems QPTs are signalled by
criticalities in the concurrence function adopted to measure
bipartite entanglement. To this end we analyze the ground-state energy of the Hamiltonian model given by equation (\ref{hamzl}) against the parameter $\tau$. Exploiting eq. (\ref{eigenvalues}) it is possible to demonstrate
that, in correspondence to the two transition points of the function $N$, the ground state energy undergoes two level crossings, which are necessary and sufficient conditions for the existence of QPTs . We may thus conclude that the results we have obtained for
the negativity $N$ in correspondence to low, but not zero, values
of $T$, stem from the existence of critical
points in the space of the parameters characterizing our system
and thus are signatures of the presence of QPTs.

It is in addition important to stress that the first
transition (the one from $N=0$ to $N= 0.5$) occurs when $2\tau = \omega$. However a positive value of $\gamma$ partially destroys the quantum correlations between the
two spins not allowing the system to reach a maximally
entangled state. Finally when $\tau \geq \omega+3\gamma$ the linear interaction can prevail on both the
quadratic interaction and the external magnetic field such that
the system can move to a maximally entangled condition. This
behavior does not depend on our choice $\gamma=\omega=1$ but manifests itself qualitatively in the same
way whatever the two parameters are fixed at.

Recent papers \cite{vedr, toth} have demonstrated that the behavior of some thermodynamical
quantities, such as for example the internal energy, the magnetic
susceptibility or the heat capacity, can reveal entanglement
between the microscopic constituents of a macroscopic sample, suggesting an intriguing approach of experimental interest to witness entanglement. We are thus stimulated to search a thermodynamical quantity of transparent physical meaning reflecting the features exhibited by the negativity function in the system parameter space as they appear in Fig.1. To this end we analyze the heat capacity , which is defined as
\begin{equation}
C_{V}=\frac{\partial U}{\partial T}
\end{equation}
where $U=T^{2}\frac{\partial \ln{Z}}{\partial T}$ is
the mean energy of the system in the thermal state
(\ref{thermalstatezl}).\\
Figure 2 reports the dependence of $C_{V}$ on $\tau$, straightforwardly got exploiting the closed form of $Z$ given by eq. (\ref{partfunctzl}), together with the negativity. The fixed value of temperature $T$ in these plots is $0.6$ and has been chosen taking into account the fact that a temperature close enough to zero would have given a flat almost zero $C_{V}$, while an high enough temperature would have destroyed thermal entanglement indeed making unsuccessful our attempt to compare negativity and heat capacity.\\
In order to realize a wider intermediate phase we fix $\gamma$ greater than one, used in Fig.\ref{negzl1}, so that the transition to the upper
phase will occur at a greater value of $\tau$.
\begin{figure}[h]
\begin{center}
\includegraphics[width=250pt]{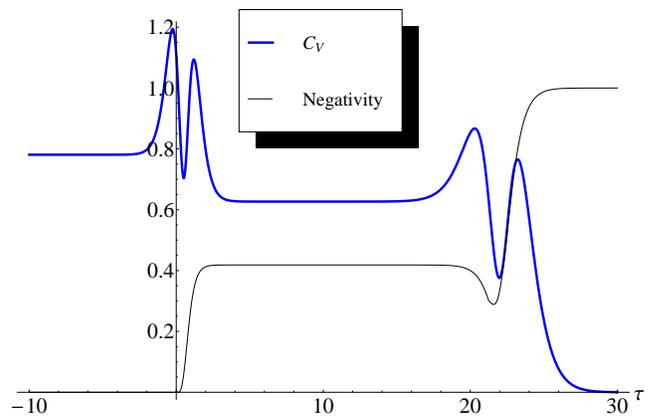}
\end{center}
\caption{Heat capacity and system negativity plotted against $\tau$ when $T=0.6$, $\gamma = 7$ and $\omega =1$}
\label{czl1}
\end{figure}
Fig.2 describes our main result demonstrating the existence of a strong similarity between the behaviors of the heat capacity of the system and the one of negativity against $\tau$. As the negativity
function, also the heat capacity is characterized by three phases
each one correspondent to a well defined value. Moreover
the higher the entanglement is, the lower is heat capacity and
viceversa. It is worth noting that such a result is a general property that is independent on the region of the parameter space ($\tau$, $\gamma$, $\omega$) here analyzed for simplicity.
\section{Discussion and conclusions}
Let us begin by providing reasonable arguments in favor of our choice of performing a comparison between negativity and heat capacity. To this end we report in Fig.3 the structure of the low-lying energy states of Hamiltonian (\ref{hamzl}).\\
\begin{figure}[h]
\begin{center}
\includegraphics[width=250pt]{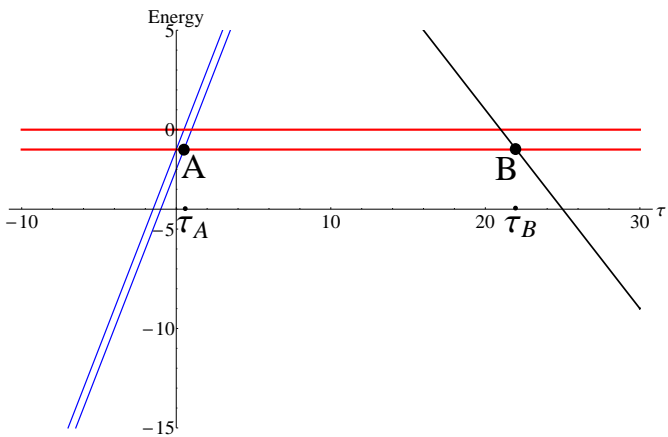}
\end{center}
\caption{Low-lying energy levels of $H$ (in units of $t$) versus $\tau$ when $\gamma = 7$ and $\omega =1$. Two crossing points, \textbf{A} and \textbf{B} in the plot, are clearly visible in correspondence to the values $\tau_{A}=\frac{1}{2}$ and $\tau_{B}=22$}
\label{czl1}
\end{figure}
The presence of points (\textbf{A} and \textbf{B}) where clearly the derivative of the ground state energy undergoes a finite jump explains why $\tau_{A}$ and $\tau_{B}$ play the roles of quantum critical points in the behavior of negativity. It is worth noting that Fig.3 shows in addition the occurrence of level crossings other than \textbf{A} and \textbf{B} around $\tau_{A}$ and $\tau_{B}$. As a consequence, the gaps between the low-lying energy levels exhibit structural changes going to zero and then rapidly increasing again. This fact suggests to test such a behavior through the heat capacity, which is by definition very sensitive to the structure of the low-lying energy levels at low temperature. The bouncing behavior of heat capacity in the vicinity of quantum critical points $\tau_{A}$ and $\tau_{B}$ as reported in Fig.2 is then easily interpreted in terms of these oscillations in the energy gaps between the accessible states of the system.
The closer these levels are, the easier is for the system itself to exchange energy with the bath and the higher heat capacity is. Far from the crossing points, where the ground state energy is analytical, these gaps are constants (some level crossing may occur between high excited states, but these ones can not affect the thermodynamics of our system as long as the temperature is low enough) and heat capacity is thus independent on $\tau$: when $\tau < \tau_{A}$ there exist two levels, one of which plays the role of ground state, whose gap is constant and smaller than the mean thermal energy at $T=0.6$. Heat capacity is thus constant and different from zero. The same can be said for the heat capacity when $\tau_{A}<\tau<\tau_{B}$, while for $\tau >\tau_{B}$ the gap between ground state and first excited level grows unbounded leading to zero heat capacity.\\
Another interesting feature related to the comparison between negativity and heat capacity is the observation that higher negativity is associated to lower thermal capacity and viceversa. It is possible to interpret this feature: in fact, the higher entanglement is,
the stronger the two atoms are correlated; these strong correlations then force the system to stay in its ground state not allowing any transition to excited states after thermal fluctuations since the atoms can not correlate to the bath for monogamy reasons \cite{hor}, thus resulting in a zero heat capacity.
On the contrary when the two spins are not correlated (factorized ground state),
nothing prevents such a transition to happen and heat capacity reaches its maximal
value since now the system is able to exchange thermal energy with the bath.
Here we freely speak about ground state properties even if our system is
in a mixed thermal state.
This is legitimated by the temperature-independence of the existence of a quantum critical point which in turn only depends on the energy spectrum properties as seen in the parameter space.\\
Moreover the temperature range we are working in (from $T\sim$nK up to $T\sim100$nK) is such that quantum critical effects are still clearly evident, showing how almost all thermal and statistical features indeed directly originate from lowest energy levels properties.\\
In conclusion, we wish to give some final remarks.\\
The novel result of this paper is reported in Fig.2 where we establish a direct conceptual link, in the parameter space, between negativity and heat capacity, giving in addition a transparent physical interpretation of it. We claim indeed that this paper for the first time successfully seeks signatures, at a thermodynamical level, of the thermal entanglement get established in a two-atom Bose-Hubbard model at low temperature, finding a surprising systematic correspondence at the critical points between quantum phase transitions and a peculiar oscillating behavior of the heat capacity. \\
In connection with the simplicity of the Hamiltonian model on which our exact results are based, we wish to emphasize that such kind of models has recently gained interest for example in the contest of the study of tunneling phenomena \cite{lin}, mostly connected to Josephson tunneling between spin condensates \cite{bosjo}. It is moreover appropriate to underline that our physical analysis, as reported in this section, relies on arguments which might be useful in the interpretation of analogous physical properties related with more complex physical scenarios (many-atom Bose-Hubbard models, central spin systems, spin chains, etc).


\begin{thebibliography}{36}

\bibitem{sac} S. Sachdev, \emph{Quantum Phase Transitions}, Cambridge University Press, 2008

\bibitem{rev} L. Amico, R. Fazio, A. Osterloh, V. Vedral, Rev. Mod. Phys. \textbf{80}, 517 (2008)

\bibitem{wa} X. Wang, Phys. Rev. A \textbf{66}, 034302 (2002)

\bibitem{expqpt} M. Greiner, O. Mandel, T. Esslinger, T.W. H\"{a}nsch, I. Bloch, Nature \textbf{415}, 39 (2002)

\bibitem{bon} L. Bonnes, S. Wessel, Phys. Rev. Lett. \textbf{106}, 185302 (2011)

\bibitem{wan} X. Wang, S.-J. Gu, J. Phys. A: Math. Theor. \textbf{40}, 10759 (2007)

\bibitem{zh} X. Zhang, C.-L. Hung, S.-K. Tung, N. Gemelke, C. Chin, New J. Phys. \textbf{13}, 045011 (2011)

\bibitem{dhar} A. Dhar, T. Mishra, R.V. Pai, B.P. Das, Phys. Rev. A \textbf{83}, 053621 (2011)

\bibitem{ost} A. Osterloh, L. Amico, G. Falci, R. Fazio, Nature \textbf{410}, 608 (2002)

\bibitem{shi} E. Shimshoni, G. Morigi, S. Fishman, Phys. Rev. Lett. \textbf{106}, 010401 (2011)

\bibitem{wu} L.-A. Wu, M.S. Sarandy, D.A. Lidar, Phys. Rev. Lett. \textbf{93}, 250404 (2004)

\bibitem{osb} T.J. Osborne, M.A. Nielsen, Phys. Rev. A \textbf{66}, 032110 (2002)

\bibitem{vid} J. Vidal, G. Palacios, R. Mosseri, Phys. Rev. A \textbf{69}, 022107 (2004)

\bibitem{dil} R. Dillenschneider, Phys. Rev. B \textbf{78}, 224413 (2008)

\bibitem{sma} P. Smacchia, L. Amico, P. Facchi, R. Fazio, G. Florio, S. Pascazio, V. Vedral, arXiv: 1105.0852 (2011)

\bibitem{or} R. Or\'{u}s, T.-C. Wei, Phys. Rev. B \textbf{82}, 155120 (2010)

\bibitem{blo} I. Bloch, J. Dalibard, W. Zwerger, Rev. Mod. Phys. \textbf{80}, 885 (2008)

\bibitem{rizzi} M. Rizzi, D. Rossini, G. De Chiara, S. Montangero, R. Fazio,  Phys. Rev. Lett. \textbf{95}, 240404 (2005)

\bibitem{bai} D. Baillie, P.B. Blakie, Phys. Rev. A \textbf{80}, 033620 (2009)

\bibitem{chn} B.-L. Chen, X.-B. Huang, S.-P. Kou, Y. Zhang, Phys. Rev. A \textbf{78}, 043603 (2008)

\bibitem{gen} S. Genway, A.F. Ho, D.K.K. Lee, Phys. Rev. Lett. \textbf{105}, 260402 (2010)

\bibitem{zz} S.Q. Zhou, D.M. Ceperley, Phys. Rev. A \textbf{81}, 013402 (2010)

\bibitem{projectors} A. Imambekov, M. Lukin, E. Demler, Phys. Rev. A \textbf{68}, 063602 (2003)

\bibitem{hameff} C.K. Law, H. Pu, N.P. Bigelow, Phys. Rev. Lett. \textbf{81}, 5257 (1998)

\bibitem{expbhqpt} W. Zwerger, J. Opt. B: Quantum Semiclass. Opt. \textbf{5}, S9 (2003)

\bibitem{longhi} S. Longhi, J. Phys. B: At. Mol. Opt. Phys. \textbf{44}, 051001 (2011)

\bibitem{zieg} K. Ziegler, Phys. Rev. A \textbf{81}, 034701 (2010)

\bibitem{lin} J. Links, A. Foerster, A.P. Tonel, G. Santos, Ann. Henri Poincar\'{e} \textbf{7}, 1591 (2006)

\bibitem{leg} A.J. Leggett, Rev. Mod. Phys. \textbf{73}, 307 (2001)

\bibitem{peres} A. Peres, Phys. Rev. Lett. \textbf{77}, 1413 (1996)

\bibitem{zhangli} G.-F. Zhang, S.-S. Li, Optics Communications \textbf{260}, 347 (2006)

\bibitem{zvy} A. A. Zvyagin, Phys. Rev. B \textbf{80}, 144408 (2009)

\bibitem{vedr} M. $\mathrm{Wie\acute{s}niak}$, V. Vedral, $\mathrm{\breve{C}}$. Brukner, Phys. Rev. B \textbf{78}, 064108 (2008)

\bibitem{toth} G. T\'{o}th, Phys. Rev. A \textbf{71}, 010301(R) (2005)

\bibitem{hor} R. Horodecki, P. Horodecki, M. Horodecki, K. Horodecki, Rev. Mod. Phys. \textbf{81}, 865 (2009)

\bibitem{bosjo} R. Gati, M.K. Oberthaler, J. Phys. B: At. Mol. Opt. Phys. \textbf{40}, R61 (2007)

\end{thebibliography}
\end{document}